# Radiometric Determination of Uranium in Natural Waters after Enrichment and Separation by Cation-Exchange and Liquid-Liquid Extraction

I. Pashalidis[1] and H. Tsertos[2][*]

1) Department of Chemistry, University of Cyprus, P.O. Box 20537, Cy-1678 Nicosia, Cyprus.
2) Department of Physics, University of Cyprus, P.O. Box 20537, Cy-1678 Nicosia, Cyprus.

The alpha-radiometric determination of uranium after its pre-concentration from natural water samples using the cation-exchange resin Chelex-100, its selective extraction by tributylphosphate and electrodeposition on stainless steel discs is reported. The validity of the separation procedure and the chemical recoveries were checked by addition of uranium standard solution as well as by tracing with $U^{232}$. The average uranium yield was determined to be $(97 \pm 2)$ % for the cation-exchange, $(95 \pm 2)$ % for the liquid-liquid extraction, and more than 99% for the electrodeposition. Employing high-resolution α-spectroscopy, the measured activity of the $U^{238}$ and $U^{234}$ radioisotopes was found to be of similar magnitude; i.e. ~7 mBq L$^{-1}$ and ~35 mBq L$^{-1}$ for ground- and seawater samples, respectively. The energy resolution (FWHM) of the α–peaks was 22 keV, while the Minimum Detectable Activity (MDA) was estimated to be 1 mBq L$^{-1}$ (at the 95% confidence limit).

[*] *Corresponding author. E-mail: tsertos@ucy.ac.cy, Fax: +357-22339060.*

**Introduction**

The determination of background concentrations and levels of environmental contamination from facilities handling enriched uranium or due to the military use of depleted uranium can be made more accurately if the isotopic ratio is measured in addition to the total uranium concentration. Furthermore, accurate knowledge of uranium isotopic ratios in natural systems (e.g. seawater) is of particular interest regarding geochronology, paleothermometry and pollution budgets. [1,2]

Several methods, such as spectrophotometry,[3] X–ray fluorescence spectroscopy,[4] inductive coupled plasma mass spectrometry,[2,5] neutron activation analysis,[6] voltametry,[7] alpha spectroscopy,[8] alpha liquid scintilation[9] etc., have been used to measure uranium concentration in natural waters. However, only mass spectrometry, neutron activation analysis and alpha spectroscopy are useful for determining isotopic ratios. Regarding alpha spectroscopy, it has the advantage of low cost and robust radioanalytical method, which can widely be used in routine analysis for the determination of $U^{234}/U^{238}$ and $U^{235}/U^{238}$ isotopic ratios in greatly enriched samples.

Pre-concentration and isolation of uranium from large amounts of the aquatic matrix is necessary because of the low uranium concentrations in natural waters and the interference of inactive substances with the alpha particles emitted, which eventually leads to lower spectral resolution and higher detection limits. Pre-concentration and isolation procedures include various pre-analytical techniques such as co-precipitation, extraction and ion-exchange.[8]

In this paper a relatively simple and effective method for the pre-concentration and separation of uranium from natural waters (e.g. seawater and groundwater) and its alpha radiometric determination after electrodeposition on stainless steel discs is presented. This method combines cation-exchange and liquid-liquid extraction techniques and is applicable to various aquatic solutions. In addition, the employment of high-resolution α–spectroscopy allows an accurate determination of the activity of the $U^{238}$ and $U^{234}$ radioisotopes, even though they appear at very low concentrations. The measurements have been carried out in the Laboratories of Radioanalytical Chemistry and Nuclear Physics of the University of Cyprus.

**Experimental**

*Reagents and instruments*

In all experiments analytical grade reagents and deionized water was used. The $Pu^{236}$-Tracer solution was obtained from AEA Technology, Nuclear Science. The Chelex-100 resin (100-200 mesh, Bio-Rad Labs) was used as received. Arsenazo(III) solutions were prepared by dissolving appropriate amounts of arsenazo(III) salt (obtained from Aldrich Co) in aquatic 0.01 M $HClO_4$. $HNO_3$ solutions of various concentrations (e.g. 2 and 8 M $HNO_3$) were prepared by diluting a concentrated solution (68% $HNO_3$). 30% v/v TBP solution was freshly prepared by mixing the appropriate volumes of tributyl phosphate (TBP) and dodecan, both obtained from Aldrich Co. $(NH_4)_2SO_4$ solutions were prepared by dissolving ammonium sulfate (99.999%, Aldrich Co) in deionized water.

Photometric and radiometric analysis of uranium was performed using a Shimadzu UV-2401 spectrophotometer and a high-resolution α-spectroscopic system equipped with ULTRA ion-implanted silicon detectors, described below.

*Sample preparation*

Sample collection included surface seawater and groundwater from three different coastal and one inland area of the island, respectively. 5-L water samples were spiked with 1.53 Bq of $Pu^{236}$ and 5 mBq of $U^{232}$ using 200 µL of a tracer solution in 2 M $HNO_3$. Isotopic Standard Reference Tracer Solution was dissolved in 2 M $HNO_3$ to prepare the spikes.

*Uranium separation by cation exchange resin*

The separation of uranium from the aqueous matrix was performed by cation exchange using the Chelex-100 resin.[10,11] 10 g of Chelex-100 resin and 20 g of ammonium acetate were added to 5-L samples contained in 5-L polypropylene beaker and stirred on a magnetic stirring table. The pH of the solutions was then adjusted to the desired value (pH = 5.3), with the aid of a pH meter, using 2 M $HNO_3$ solution. After pH adjustment the beakers were sealed and stirred for at least an hour. The resin was separated from the solution by filtration, with the aid of a Buchner funnel (with sintered glass disc) and washed with 20 mL of distilled water.

*Uranium extraction and electrodeposition*

Following this initial separation, the uranium was eluted from the Chelex-100 resin with 30 mL of 2 M $HNO_3$ solution. The acidic solutions were evaporated to incipient dryness and the residual was dissolved in 15 mL of 8 M $HNO_3$ solution. The latter and 15 mL of the 30% TBP solution were added in a 100 mL, short stemmed, separatory funnel, and the content was shaken vigorously for about 30 seconds. This extraction procedure was repeated once more and the aqueous phase was discarded. The uranium was removed from the organic phase by extracting successively two times with 30 mL of distilled water and collecting the aqueous phase in a 150 mL beaker.

The aqueous solutions were evaporated to incipient dryness and the residual was dissolved in 10 mL of 0.15 M $(NH_4)_2SO_4$ solution. From this solution uranium was directly electrodeposited on a stainless steel disc (diameter of the active area 10 mm) at a voltage of 15 V and a current of 0.4 A for 2 hours.[12]

*High-resolution α–spectroscopy*

A stand-alone high-resolution spectroscopic system is used for the measurement of the energy spectrum of the emitted α–particles in the energy range between 3000 ke*V* and 8000 *keV*. The system consists of cylindrical ULTRA ion-implanted silicon detectors (EG&G ORTEC BU-020-450-300) that are characterized by an active area of 450 $mm^2$ and by a depletion depth of 300 μm. The detectors are mounted on the inside of a low-background vacuum chamber with low-background sample holders (EG&G ORTEC 808). The spectroscopic system is linked with a Multi-Channel Buffer (MCB) which is a PC-based plug-in PCI card consisting of an 2k Analogue-to-Digital Converter (ADC). An advanced Multi-Channel Analyser (MCA) emulation software (MAESTRO-32) enables data acquisition, storage, display and online analysis of the acquired α-spectra.

The energy resolution and detection efficiency were determined using a calibrated mixed open source[†], consisting of $Np^{237}$, $Am^{241}$, and $Cm^{244}$ at nominal activities of 150 Bq, 100 Bq, and 100 Bq, respectively. The source material has been uniformly electrodeposited on a stainless steel disc to an effective diameter of 10 mm. The source disc was mounted on a sample holder at a distance of about 10 cm from the detector. A measured α–spectrum from this source is shown in Fig. 1.

---

[†] Manufactured by AEA Technology QSA GmbH, Germany and calibrated by the German calibration laboratory for measurements of radioactivity (DKD).

The energy resolution (FWHM) achieved in the measurements was 22 keV, whereas the detection efficiency for each detector was calculated to be 0.35% of 4π. Considering an uncertainty of 3% in the source activity, the mean uncertainty in the calculated efficiency was estimated to be about 5%. Finally, the Minimum Detectable Activity (MDA) reached in the measurements was estimated to be about 1 mBq per liter (mBq $L^{-1}$) at the 95% confidence limit.

The offline analysis of each measured α–ray spectrum has been carried out by a dedicated software programme (AlphaVision-32)[13], which performs a simultaneous fit to all the statistically significant peaks appearing in the spectrum. Reports were generated which included the centroid channel, energy, net area counts, background counts, intensity and width of identified and unidentified peaks in the spectrum, as well as the calculated activity in $Bq\ L^{-1}$ for each detected radionuclide. It is noted here that prior to the samples measurement, the background was carefully measured under identical measurement conditions and was found to be about 24 counts per day within the energy range of 3–8 MeV. It has been subtracted from the measured α–ray spectra of each sample.

**Results and Discussion**

Uranium in surface seawater exists quantitatively in the hexavalent oxidation state and the anionic uranium-carbonato complexes, $UO_2(CO_3)_2^{2-}$ and $UO_2(CO_3)_3^{4-}$, govern the chemical behaviour of uranium in solution under seawater conditions.[1,2] Acidification (pH < 7) of seawater samples results in the destabilisation of the carbonato species and the formation of uranium(VI)-hydroxo- or uranium(VI)-aquo- complexes.[14] The separation and pre-concentration of uranium by the cation exchange resin Chelex-100 was performed at pH = 5.3, because at this pH value the cationic uranium(VI) species, $UO_2^{2+}$ and $UO_2OH^+$, are predominant in solution and the competition of protons, regarding cation-exchange, relatively low. The uranium recovery during the cation exchange procedure was followed by spectrophotometry using arsenazo(III)[3] and was calculated to be (97 ± 2) %. Determination of the uranium yield after the various pre-analytical steps (e.g. cation-exchange and liquid-liquid extraction) was performed by photometry using arsenazo III, because the alpha radiometric determination by means of surface barrier detectors was not possible or resulted in very bad quality of spectra. Only the total uranium yield could be determined alpha-radiometrically by tracing the samples with a $Pu^{236}/U^{232}$ standard tracer solution.

After cation-exchange, the samples were largely free of extraneous ions, but contained still too much heavy metal ions, which disturbed the homogeneous electrodeposition of uranium on the stainless steel discs. Therefore, the samples were purified further by a standard tributyl-phosphate extraction procedure,[15] which removes almost all remaining interferences. The uranium yield after the extraction procedures was determined by photometry and was calculated to be (93 ± 2) %. Hence, the overall chemical recovery of uranium was determined to be (90 ± 3) %.

The electrodeposition of uranium on stainless steel discs was optimised and resulted in excellent yields, generally over 99%. In order to check the reproducibility of the electrodeposition procedure and also the detection efficiency and performance of the α−spectroscopic system, six identical samples were prepared from the $Pu^{236}$ tracer solution with an initial activity of 1 Bq each. After electrodeposition to stainless steel discs (10 mm in diameter) and measuring the corresponding α−spectra (the $Pu^{236}$ main

α−peak appears at an energy of ~5770 keV), the initial activity of 1 Bq could be reproduced within 5% by all six samples.

The measured α−ray spectra from Cyprus natural water samples have revealed rather low activity concentration of $U^{238}$ and $U^{234}$ radioisotopes. Therefore, in order to obtain the α−peaks with sufficient statistics (5-10% statistical error), the accumulation time had to be rather long (about 7−10 days for each sample). Nevertheless, the α−spectra were obtained at high-energy resolution (FWHM ~22 keV) and nearly background-free, demonstrating the effectiveness and specificity of the procedure as well as the excellent long-term stability of the α−spectroscopic system. A typical example of the quality of the measured α−spectra is presented in Fig. 2 for a seawater sample. As can be seen, the two main α−peaks due to the $U^{238}$ and $U^{234}$ radioisotopes (with energies of ~4198 keV and of ~4776 keV, respectively) revealed almost comparable intensities. No other statistically significant α−peaks than those of the $U^{238}$ and $U^{234}$ radioisotopes were detected in the measured samples. The small contribution (~2%) due to the $U^{235}$ radioisotope was below the experimental detection limit.

The activity concentration, calculated from the spectrum shown in Fig. 2, was (35.0 ± 1.9) mBq $L^{-1}$ and (37.5 ± 2.0) mBq $L^{-1}$ for the $U^{238}$ and $U^{234}$ radioisotopes, respectively. The activity ratio ($U^{234}/U^{238}$) was then estimated to be 1.07 ± 0.08. The other seawater samples that were investigated exhibited very similar activity concentrations. Uranium concentration in groundwater revealed even lower values, which amounted to (6.6 ± 0.8) mBq $L^{-1}$ and (6.7 ± 0.8) mBq $L^{-1}$ for the $U^{238}$ and $U^{234}$ radioisotopes, respectively. The corresponding activity ratio ($U^{234}/U^{238}$) was calculated to be 1.01 ± 0.15.

In naturally occurring uranium radioisotopes, the activity ratio $A(U^{234}/U^{238})$ can be calculated according to the following expression:

$$A(U^{234}/U^{238}) = \frac{T_{U^{238}}}{T_{U^{234}}} \cdot \frac{f_{U^{234}}}{f_{U^{238}}} \cdot \frac{M_{U^{238}}}{M_{U^{234}}} \quad (1)$$

where $T$ is the half-life, $f$ is the fractional atomic abundance, and $M$ is the atomic mass of the corresponding radioisotope indicated. Taking current values from the literature,[16] the above formula yields $A(U^{234}/U^{238}) \cong 1.05$.

Within the experimental uncertainties quoted, the measured activity ratios ($U^{234}/U^{238}$) fit very well to the expected value of ~1.05 (Eq. (1)), revealing the natural origin of the measured uranium radioisotopes.

**Conclusions**

Exploitation of a rather robust and effective radioanalytical method for the pre-concentration and separation of uranium from natural waters (e.g. seawater and groundwater samples) as well as of high-resolution α–spectroscopy provide a sensitive experimental tool in measuring activity concentration of the $U^{238}$ and $U^{234}$ radioisotopes and determining their activity ratio after electrodeposition on stainless steel discs. In general, the measured α–spectra from ground- and seawater samples studied were obtained almost background-free and revealed rather low activity concentration of the $U^{238}$ and $U^{234}$ radioisotopes, exhibiting values of ~7 mBq L$^{-1}$ and ~35 mBq L$^{-1}$ for ground- and seawater samples, respectively. In all samples studied, the activity ($U^{234}/U^{238}$) ratio was found to be consistent with the value of ~1.05, which is expected from naturally occurring uranium.

**Acknowledgements**

This work is conducted with financial support from the Cyprus Research Promotion Foundation (Grant No. 45/2001) and partially by the University of Cyprus.

**FIGURE CAPTIONS**

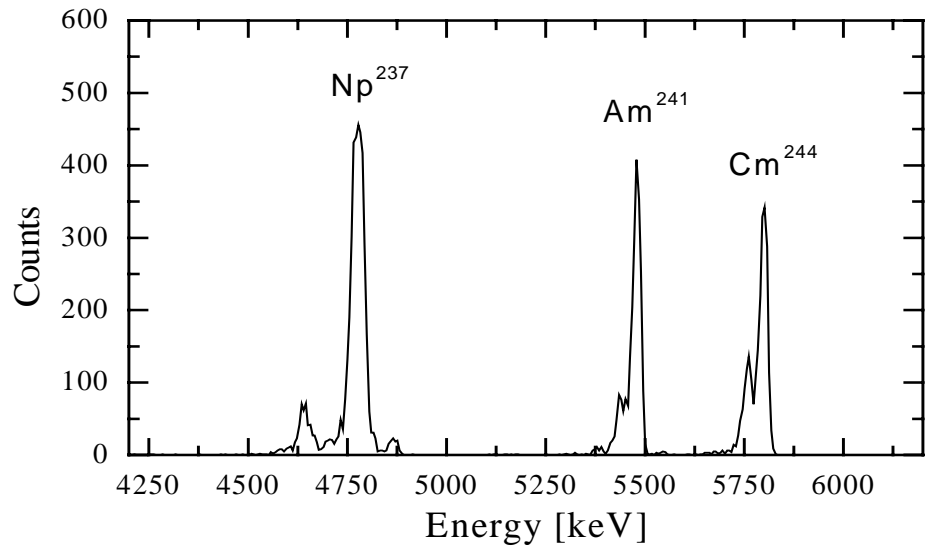

**Figure 1**. Alpha-ray spectrum measured from a calibrated mixed open source ($Np^{237}$, $Am^{241}$ and $Cm^{244}$). The source material has been uniformly electrodeposited on a stainless steel holder disc to an effective diameter of 10 mm. The high-resolution α-detector had an active area of 450 $mm^2$ and was mounted to a distance of ~10 cm from the source. The accumulation time was 7,200 *s* (2 h).

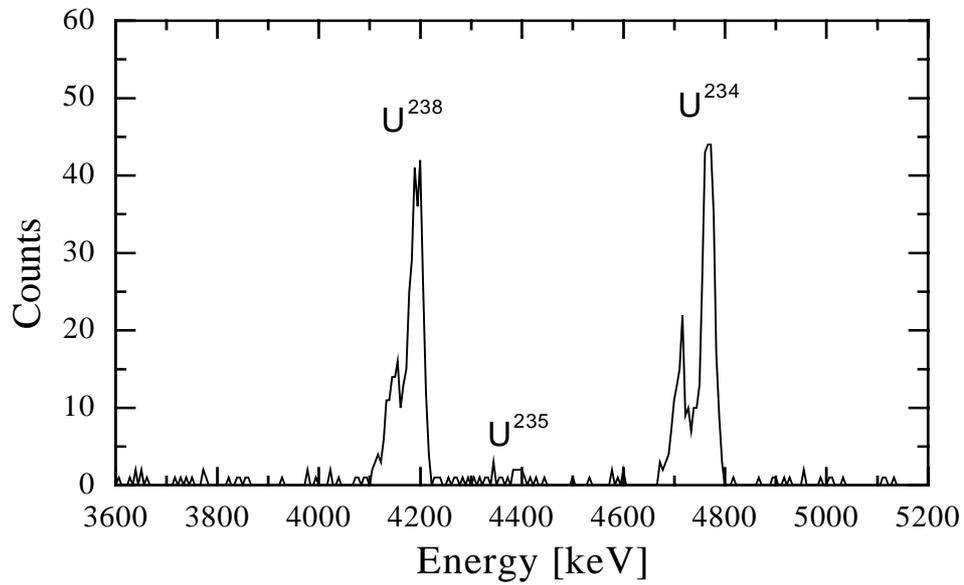

**Figure 2**. Alpha-ray spectrum measured from a natural seawater sample, prepared as described in the text. The $U^{238}$ and $U^{234}$ α-peaks have been observed at similar intensities and nearly background-free. The α-detector and the geometry used were the same as those given in Fig. 1. The accumulation time was 166 h (6.9 days).